\documentclass[twocolumn,aps,pra,superscriptaddress]{revtex4-2}

\usepackage{times}
\usepackage{graphicx}
\usepackage{epsfig}
\usepackage{amsfonts}
\usepackage{amsmath}
\usepackage{amssymb}
\usepackage{color}
\usepackage{multirow}
\usepackage[colorlinks=true,citecolor=blue,urlcolor=blue]{hyperref}

\def\endproof{\vrule height6pt width6pt depth0pt}

\newtheorem{theorem}{Theorem}
\newtheorem{conjecture}[theorem]{Conjecture}
\newtheorem{definition}{Definition}
\newtheorem{lemma}[theorem]{Lemma}
\newtheorem{corollary}[theorem]{Corollary}

\begin{document}

%lucaseaporto@gmail.com,nufalas@gmail.com,tcunha@unicamp.br,adan@us.es

%%%%%%%%%%%%%%%%%%%%%%%%%%%%%%%%%%%%%%%%%%%%%%%%%%%%%%%%%%%%%%%%%%%

\title{The quantum maxima for the basic graphs of exclusivity are not reachable in Bell scenarios}

%%%%%%%%%%%%%%%%%%%%%%%%%%%%%%%%%%%%%%%%%%%%%%%%%%%%%%%%%%%%%%%%%%%

\author{Lucas E. A. Porto}
\affiliation{Instituto de F\'{i}sica Gleb Wataghin, Universidade Estadual de Campinas (Unicamp),
Rua S\'{e}rgio Buarque de Holanda 777, Campinas, S\~{a}o Paulo 13083-859, Brazil}

\author{Rafael Rabelo}
\affiliation{Instituto de F\'{i}sica Gleb Wataghin, Universidade Estadual de Campinas (Unicamp),
Rua S\'{e}rgio Buarque de Holanda 777, Campinas, S\~{a}o Paulo 13083-859, Brazil}

\author{Marcelo Terra Cunha}
\affiliation{Instituto de Matem\'{a}tica, Estat\'{i}stica e Computa\c{c}\~{a}o Cient\'{i}fica, Universidade Estadual de Campinas (Unicamp), Rua S\'{e}rgio Buarque de Holanda 651, Campinas, S\~{a}o Paulo 13083-859, Brazil}

\author{Ad\'{a}n~Cabello}
\affiliation{Departamento de F\'{\i}sica Aplicada II, Universidad de Sevilla, E-41012 Sevilla, Spain}
\affiliation{Instituto Carlos~I de F\'{\i}sica Te\'orica y Computacional, Universidad de Sevilla, E-41012 Sevilla, Spain}

%%%%%%%%%%%%%%%%%%%%%%%%%%%%%%%%%%%%%%%%%%%%%%%%%%%%%%%%%%%%%%%%%%%

\begin{abstract}
A necessary condition for the probabilities of a set of events to exhibit Bell nonlocality or Kochen-Specker contextuality is that the graph of exclusivity of the events contains induced odd cycles with five or more vertices, called odd holes, or their complements, called odd antiholes. From this perspective, events whose graph of exclusivity are odd holes or antiholes are the building blocks of contextuality. For any odd hole or antihole, any assignment of probabilities allowed by quantum theory can be achieved in specific contextuality scenarios. However, here we prove that, for any odd hole, the probabilities that attain the quantum maxima cannot be achieved in Bell scenarios. We also prove it for the simplest odd antiholes. This leads us to the conjecture that the quantum maxima for any of the building blocks cannot be achieved in Bell scenarios. This result sheds light on why the problem of whether a probability assignment is quantum is decidable, while whether a probability assignment within a given Bell scenario is quantum is, in general, undecidable. This also helps to understand why identifying principles for quantum correlations is simpler when we start by identifying principles for quantum sets of probabilities defined with no reference to specific scenarios.
\end{abstract}

%%%%%%%%%%%%%%%%%%%%%%%%%%%%%%%%%%%%%%%%%%%%%%%%%%%%%%%%%%%%%%%%%%%

\maketitle

%%%%%%%%%%%%%%%%%%%%%%%%%%%%%%%%%%%%%%%%%%%%%%%%%%%%%%%%%%%%%%%%%%%

\section{Motivation}

%%%%%%%%%%%%%%%%%%%%%%%%%%%%%%%%%%%%%%%%%%%%%%%%%%%%%%%%%%%%%%%%%%%

Bell nonlocality \cite{BCPSW14}, i.e., the existence of correlations between outcomes of space-like separated measurements which are impossible to achieve with local hidden variable models (because there is no joint probability distribution that reproduces the probability distributions of all contexts), is one of the most fascinating properties of nature. However, among the many possible ways of being Bell nonlocal, only some of them occur in nature: those allowed by quantum theory. 
A fundamental question is: is there any principle that selects these ways? \cite{PR94}.
This question is related to a more profound question: how come the quantum? \cite{Wheeler86}.

So far, all proposed principles for quantum nonlocal correlations, including no-signalling \cite{PR94}, information causality \cite{PPKSWZ09}, macroscopic locality \cite{NW09}, and local orthogonality \cite{FSABCLA13}, have failed to identify the set of quantum correlations, even for the simplest Bell scenario \cite{SGAN18}. However, one principle, called the totalitarian principle \cite{Cabello19b} or the principle of plenitude \cite{BCGKL22}, is able to select the quantum correlations for all the scenarios where Bell nonlocality or Kochen-Specker (KS) contextuality can happen, under the assumption that the observables are ideal \cite{Cabello19a}. Ideal observables are those that, they and all their coarse–grained versions, can be measured ideally (i.e., giving the same outcome when the measurement is repeated, and without disturbing any compatible observable). Hereafter, we will refer to the contextuality produced by ideal observables as KS contextuality \cite{BCGKL22}. 

Every quantum Bell nonlocal correlation can be achieved with ideal observables \cite{Cabello19a}. Then, how is that we have a principle for the sets of quantum correlations for ideal observables but not for the sets of quantum nonlocal correlations? Arguably, the reason is that it is difficult to understand nonlocal correlations by only looking at Bell scenarios, because, as we shall show, they add constraints that make the mathematical characterisation of the sets of quantum correlations much more difficult than it is in other scenarios. The aim of this paper is to prove some results that illustrate this point. Specifically, we show that the most basic ingredients for KS contextuality cannot reach their maxima in any Bell scenario.

The structure of the paper is the following. In Sec.~\ref{tools} we collect some concepts and results that will be used later. In Sec.~\ref{results} we state our results. In Sec.~\ref{conclusions}, we discuss how they shed light on why it is so difficult to understand quantum Bell nonlocal correlations by only discussing Bell scenarios. We also list some open problems. The proofs are presented in Sec.~\ref{proofs}. 

%%%%%%%%%%%%%%%%%%%%%%%%%%%%%%%%%%%%%%%%%%%%%%%%%%%%%%%%%%%%%%%%%%%

\section{Concepts and previous results}
\label{tools}

%%%%%%%%%%%%%%%%%%%%%%%%%%%%%%%%%%%%%%%%%%%%%%%%%%%%%%%%%%%%%%%%%%%

Here we recall some concepts and results of the graph-theoretic approach to correlations \cite{CSW10,CSW14,RDLTC14,AT18} that are required for the main result. Readers familiar with this approach can skip this section. 

In a nutshell, this approach associates a graph to the Bell nonlocality or KS contextuality witness of every Bell or noncontextuality inequality. By studying the properties of this graph, one can obtain, e.g., the classical, quantum, and more general bounds to the witness. 

To set the ground, we consider an {\em experimental scenario}, which is defined by a set of observables, their relations of compatibility (joint measurability), and a set of possible outcomes for each observable. We call an {\em event} a statement of the form `the outcomes $a,\ldots, c$ are obtained, when the observables $x,\ldots, z$ are, respectively, jointly measured.' We will denote this event as $a,\ldots , c|x, \ldots , z$. Two events are said to be {\em mutually exclusive} if both include the same measurement but with different outcomes.

\begin{definition}
Given a set of events $\{e_i\}_{i=1}^n$, their {\em graph of exclusivity} is the $n$-vertex graph $G$ in which each event corresponds to a vertex and mutually exclusive events are represented by adjacent vertices.
\end{definition}

\begin{definition}
A {\em vertex-weighted graph} $(G,w)$ is a graph $G$ with vertex set $V(G)$ endowed with a weight assignment $w:V(G) \rightarrow \mathbb{R}_+$. 
\end{definition}

Every Bell and noncontextuality inequality can be expressed as an upper bound $\Omega$ of a positive linear combination $S$ of probabilities of events, $S = \sum_i w_i P(e_i) \leq \Omega$, with $w_i > 0$. $S$ is called a {\em witness} (of Bell nonlocality or KS contextuality, respectively).
The connection to properties of a graph comes through the vertex-weighted graph of exclusivity $(G, w)$ of the events $e_i$ in $S$ with weights $w_i$. 
Allowing non-unit weights is important for this formalism to be applicable to general witnesses. However, hereafter, we will focus on 
witnesses in which
$w_i=1$ for all $i$ and thus on standard (unweighted) graphs. This case includes many important Bell and KS contextuality witnesses such as the ones in the Clauser-Horne-Shimony-Holt Bell inequality and the Klyachko-Can-Binicio\u{g}lu-Shumovsky noncontextuality inequality \cite{CDLP13}.

The maximum value of $S$ for probability assignments consistent with the graph of exclusivity $G$ depends on which type of assignments we consider.

In a \textit{deterministic} assignment, each $e_i$ either occurs with certainty or 
does not occur, i.e., $P(e_i)=1$ or $P(e_i)=0$. We call a probability assignment \textit{classical} if it can be written as a convex combination of deterministic assignments. The maximum of $S$ for classical assignments is the \textit{independence number} of $G$.

\begin{definition}
The {\em independence number} of $G$, denoted by $\alpha(G)$, is the cardinality of the largest independent set of $G$.
A set of vertices of $G$ is {\em independent} if all the vertices in it are pairwise nonadjacent. 
\end{definition}

{\em Quantum} assignments are related to the concept of an {\em orthogonal projective representation} of $G$, while the maximum of $S$ is related to the \textit{Lovász number} of $G$.

\begin{definition} 
An {\em orthogonal projective representation} (OPR) of a graph $G$ is an assignment of a projector $\Pi_i$ (not necessarily of rank-one) onto a subspace of a $d$-dimensional Hilbert space ${\cal H}$ to each $i \in V(G)$, such that $\Pi_i \Pi_j = 0 = \Pi_j \Pi_i$ (i.e., the subspaces onto which $\Pi_i$ and $\Pi_j$ project are orthogonal), for all pairs $i,j$ of vertices that are adjacent in $G$.
\end{definition}

\begin{definition}
The {\em Lov\'asz number} of $G$ is
\begin{equation}
\vartheta(G):= \sup \sum_{i\in V(G)} \langle \psi |\Pi_i | \psi \rangle ,
\label{Lovasz}
\end{equation}
where the supremum is taken over all OPRs $\{\Pi_i\}$ of $G$ and unit vectors $|\psi \rangle$ (each of them is called a {\em handle} of the corresponding OPR) in any (finite or infinite) dimension. An OPR of $G$ is {\em Lovász-optimal} if it achieves $\vartheta(G)$ with some handle.
\end{definition}

With these definitions, the above discussion can be formalised as the following theorem, proven in Ref.~\cite{CSW14}.

\begin{theorem}
Given a sum of probabilities of events $S$ corresponding to a Bell or a noncontextuality inequality, the maximum value of $S$ achievable in classical theories (local/noncontextual models) and the maximum value of $S$ achievable in quantum theory satisfy
\begin{equation}
S \stackrel{\mbox{\tiny{L / NC}}}{\leq} \alpha(G) \stackrel{\mbox{\tiny{{\em Q}}}}{\leq} \vartheta(G),
\end{equation}
where $G$ is the graph of exclusivity associated to $S$, $\alpha(G)$ is the independence number of $G$, and $\vartheta(G)$ is the Lov\'asz number of $G$. In some scenarios, $\vartheta(G)$ might be only an upper bound to the maximum quantum value of $S$.
\label{thm:CSW_bounds}
\end{theorem}

For the purposes of this paper, a particular detail of the previous theorem deserves additional attention. This is the fact that, for some scenarios, the Lovász number of the graph of exclusivity associated to a certain inequality might not be the exact maximum value of the inequality achievable in quantum theory, but only an upper bound of it. This may occur because, in a particular scenario, there might be restrictions which are not captured by the graph of exclusivity. 

For example, in the case of Bell scenarios, measurements are local measurements on spatially separated subsystems. This implies the restriction that the projectors of the OPR must be tensor products such that each of the factors is a projector associated with a certain subsystem. Moreover, given this tensor product structure, if two events are mutually exclusive because a local measurement on subsystem $A$ has different results in each event, then another restriction is that the tensor products' factors associated to $A$ must be orthogonal.

In fact, in a Bell scenario there are different types of relations of exclusivity. For example, in a bipartite Bell scenario, two events can be mutually exclusive because a measurement on $A$ has different results in each event, because of measurement on a different subsystem $B$ has different results, or because both things occur simultaneously. To account for this, in this paper we will make use of the edge-coloured multigraph approach proposed in Ref.~\cite{RDLTC14}.

\begin{definition}
A {\em multigraph} $\Gamma=(V,E)$ is a graph with vertex set $V$ and edge set $E$ such that multiple edges between two vertices are allowed.
\end{definition}

In particular, we will be interested in a special type of multigraphs that represent the relations of exclusivity between events in $N$-partite Bell scenarios: $N$-colour edge-coloured multigraphs $\textsf{G}=(V,E)$ composed of $N$ simple graphs $G_A = (V, E_A),\ldots, G_N = (V, E_N)$ that have a common vertex set $V$ and mutually disjoint edge sets $E_A,\ldots, E_N$, such that $E = E_A \sqcup \ldots \sqcup E_N$ (where $\sqcup$ stands for disjoint union) and each $E_j$ is of a different colour. That is, we will focus on multigraphs $\textsf{G}$ that can be factorised into $N$ simple subgraphs $G_A,\ldots, G_N$, called {\em factors}, each of which spans the entire set of vertices of $\textsf{G}$, and such that all together collectively exhaust the set of edges of $\textsf{G}$. The factor $G_J$ represents the exclusivity relations associated to party (subsystem) $J$. Throughout the paper, we will refer to these multigraphs as edge-coloured (or simply coloured) multigraphs. For examples illustrating these ideas, we refer the reader to Refs.~\cite{RDLTC14, VTC22}.

The extra structure of edge-coloured multigraphs will allow us to formalise the previous discussion concerning the more restrictive nature of the quantum probability assignments in Bell scenarios thanks to the following definition.

\begin{definition}
An {\em orthogonal projective representation} (OPR) of a multigraph $\textsf{G} = (V, E)$ composed of simple graphs $G_A = (V, E_A),\ldots, G_N = (V, E_N)$ is an assignment of a projector $\Pi_i = \Pi^{A}_{i} \otimes \cdots \otimes \Pi^{N}_{i}$, where $\otimes$ denotes tensor product, to each $i \in V$ such that $\{ \Pi^{J}_{i} : i \in V \}$ constitutes an orthogonal projective representation of $G_J$, for all parties (subsystems) $J$.
\end{definition}

Notice that an OPR of an edge-coloured multigraph $\textsf{G}$ also is an OPR of the graph constructed from $\textsf{G}$ by merging all the edges connecting each two vertices into a single (colourless) one. We refer to this simple graph as the \textit{shadow} of the multigraph $\textsf{G}$. Therefore, the general question addressed in this work can be stated as follows: given a graph $G$, is there an OPR of an edge-coloured version of $G$ which is able to achieve its Lovász number, thus being a Lovász-optimal OPR of the shadow $G$? In other words, can the quantum maximum for a graph of exclusivity $G$ be achieved in a Bell scenario?

The answer to this question depends on the specific graph $G$ under consideration. There are known examples illustrating both possibilities. The Clauser-Horne-Shimony-Holt Bell inequality, for instance, achieves its maximum quantum value in a Bell scenario. On the other hand, none of the edge-colourings of the pentagon reaches its colourless Lovász number \cite{SBBC13, RDLTC14}. In this work, we analyse this question for a special class of graphs which includes the pentagon, the odd cycles with five or more vertices, and their complements (see Definition \ref{def:complement}). Except for the pentagon, it remained unclear whether the quantum maxima for them could be achieved in Bell scenarios.

The odd cycles with five or more vertices, called odd {\em holes}, and denoted by $C_n$, and their complements, called odd {\em antiholes} $\overline{C}_n$, are quite important in the study of nonclassical correlations, in the light of the perfect graph theorem.

\begin{definition}
A subgraph $H$ of a graph $G$ is an {\em induced subgraph} if, for any pair of vertices $i,j$ of $H$, $(i,j)$ is an edge of $H$ if and only if $(i,j)$ is an edge of $G$. 
\end{definition}

\begin{definition}
The {\em complement of a graph} $G$, denoted by $\overline{G}$, is a graph with the same vertex set as $G$ and in which two different vertices are adjacent if and only if they are not adjacent in $G$.
\label{def:complement}
\end{definition}

\begin{theorem}\cite{CSW14, Chudnovsky2006}
(The perfect graph theorem) If a graph $G$ has no odd cycle with five or more vertices, nor their complements, as induced subgraphs, then $\alpha(G) = \vartheta(G)$.
\label{thm:building_blocks}
\end{theorem}

In other words, if quantum theory violates a Bell or noncontextuality inequality, the graph of exclusivity of its associated witness must necessarily contain an induced odd hole or an induced odd antihole. In this sense, odd holes and odd antiholes of exclusivity constitute the building blocks of nonclassical correlations. 

%%%%%%%%%%%%%%%%%%%%%%%%%%%%%%%%%%%%%%%%%%%%%%%%%%%%%%%%%%%%%%%%%%%

\section{Results}
\label{results}

%%%%%%%%%%%%%%%%%%%%%%%%%%%%%%%%%%%%%%%%%%%%%%%%%%%%%%%%%%%%%%%%%%%

\subsection{Quantum maxima for holes in Bell scenarios}

%%%%%%%%%%%%%%%%%%%%%%%%%%%%%%%%%%%%%%%%%%%%%%%%%%%%%%%%%%%%%%%%%%%

The colourless cycles have been extensively studied in literature. Indeed, we know an analytical expression for $\vartheta(C_n)$ for any odd $n \geq 5$ \cite{Pearle70,BC90,AQBTC13}. Furthermore, the Lovász-optimal OPRs of these cycles are very well characterised, and must satisfy some strong symmetry conditions, a result which comes from self-testing \cite{Bharti2019}. With this in mind, we can prove the following theorem.

\begin{theorem}
\label{thm:NocolouredCyle}
There is no orthogonal projective representation of any coloured odd $n$-cycle, $\textsf{C}_n$, with $n \geq 5$, using at least two colours, which is a Lovász-optimal representation of the shadow $C_n$. 
\end{theorem}

In other words, this can be expressed as the following corollary.

\begin{corollary}
An odd hole cannot reach its quantum maximum in any Bell scenario.
\label{thm:NoMaxBellHole}
\end{corollary} 

This was known to be the case for $n = 5$ \cite{SBBC13}. However, it was an open problem for the other holes. This is the main contribution of this paper.

Notice that Theorem \ref{thm:NocolouredCyle} holds for any edge-colouring of the odd holes. However, it is not true that every colouring can represent events of a Bell scenario. To continue our discussion, we first need to address this point.

%%%%%%%%%%%%%%%%%%%%%%%%%%%%%%%%%%%%%%%%%%%%%%%%%%%%%%%%%%%%%%%%%%%

\subsection{When does a coloured multigraph represent exclusivity relations between events in a Bell scenario?}

%%%%%%%%%%%%%%%%%%%%%%%%%%%%%%%%%%%%%%%%%%%%%%%%%%%%%%%%%%%%%%%%%%%

The following theorem completely characterises which kind of edge-coloured multigraph can represent events in a Bell scenario.

\begin{theorem} 
\label{thm:BellStructure}
A coloured multigraph represents the relations of exclusivity between events in a Bell scenario if and only if the graph of each colour, $c$, is the disjoint union of $m_c$ complete $o_{j_c}$-partite graphs, with $j_c = 1,\ldots, m_c$. Each colour $c$ represents a party, $m_c$ is the number of measurements of the party represented by colour $c$, and $o_{j_c}$ is the number of outputs of this party's $j_c$-th measurement.
\end{theorem}

\begin{definition}
A graph $G=(V,E)$ is a complete $k$-partite graph if $V$ is the disjoint union of $k$ non-empty sets $V_k$, and two vertices, $x,y$, are adjacent if and only if $x \in V_k$ and $y\in V_l$ for $k \neq l$.
\end{definition}

%%%%%%%%%%%%%%%%%%%%%%%%%%%%%%%%%%%%%%%%%%%%%%%%%%%%%%%%%%%%%%%%%%%
% Fig. 1
%%%%%%%%%%%%%%%%%%%%%%%%%%%%%%%%%%%%%%%%%%%%%%%%%%%%%%%%%%%%%%%%%%%

\begin{figure}
\hspace{-2mm}
\includegraphics[width=8.6cm]{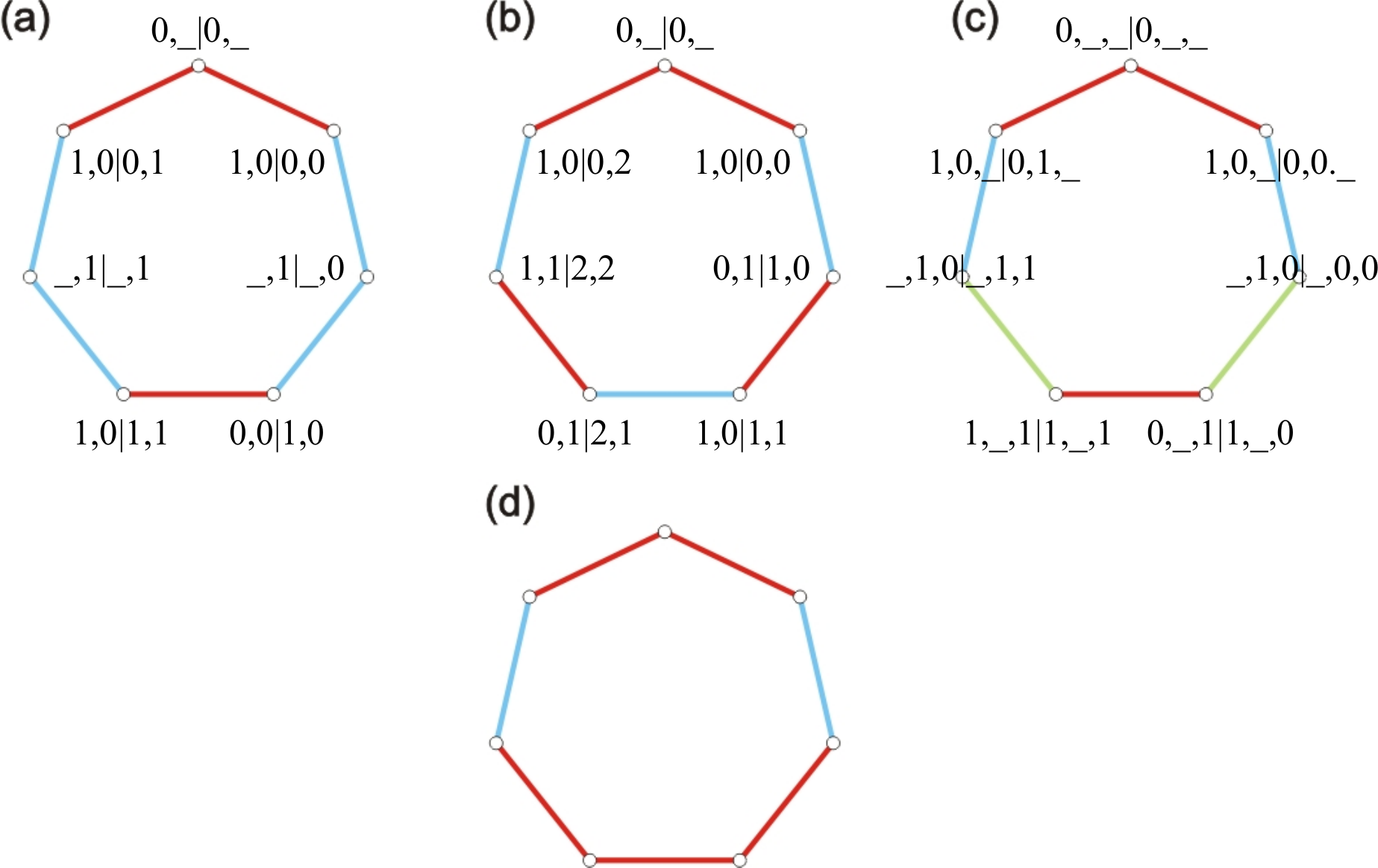}
\caption{(a), (b), and (c) are coloured graphs that represent the relations of exclusivity of events in, respectively, the $(2,2,2)$, $(2,3,2)$, and $(3,2,2)$ Bell scenarios, where $(p,m,o)$ indicates that there are $p$ parties, each with $m$ measurements, of $o$ outcomes. 
In contrast, (d) does not represent events in any Bell scenario, due to the red path of size $3$. $a,b|x,y$ denotes the event in which Alice measures $x$ and obtains $a$, and Bob measures $y$ and obtains $b$. Similarly, $a,b,c|x,y,z$ denotes the event in which Alice measures $x$ and obtains $a$, Bob measures $y$ and obtains $b$, and Charlie measures $z$ and obtains $c$. The ``\_'' on the labelings of some events means that the respective measurement is ignored.}
\label{theorem1}
\end{figure}

%%%%%%%%%%%%%%%%%%%%%%%%%%%%%%%%%%%%%%%%%%%%%%%%%%%%%%%%%%%%%%%%%%%

For example, Fig.~\ref{theorem1} presents four coloured graphs. 
Theorem~\ref{thm:BellStructure} allows us to rapidly conclude that only three of them can represent events in Bell scenarios.

Complete $k$-partite graphs can be recognised in polynomial time even when the partition into $k$ disjoint sets (see previous definition) is not supplied. Recall that complete $k$-partite graphs are equivalent to $\bar{P_3}$-free graphs, the graphs such that $\bar{P}_{3}$ (the graph with 3 vertices and a single edge) is a forbidden subgraph. Recognition of such graphs can be done in polynomial time by means of finite forbidden subgraph characterisation. Therefore, Theorem~\ref{thm:BellStructure} provides a useful way of determining whether or not a given coloured multigraph can represent events in Bell scenarios.

Moreover, it helps us to identify the simplest Bell scenario in which the coloured multigraph fits in and to find a libelling for the vertices. 
This is so because each component of each colour corresponds to a measurement on the respective part. The simplest building block would be a complete graph $K_o$ with the respective number of outputs.
Since the same local pair measurement-result commonly appears in more than one term of the Bell inequality, this gives rise to the classification in terms of complete $o$-partite graphs: each (maximal) independent set in this (connected) graph corresponds to a different outcome for the measurement specified by this component.
Finally, this naturally defines the \emph{minimal} Bell scenario in which such inequality fits.

%%%%%%%%%%%%%%%%%%%%%%%%%%%%%%%%%%%%%%%%%%%%%%%%%%%%%%%%%%%%%%%%%%%

\subsection{Quantum maxima for holes in bipartite Bell scenarios}

%%%%%%%%%%%%%%%%%%%%%%%%%%%%%%%%%%%%%%%%%%%%%%%%%%%%%%%%%%%%%%%%%%%

With the above characterisation, we can compute what is the quantum maxima for odd holes in bipartite Bell scenarios. Notice that Theorem \ref{thm:NocolouredCyle}, although valid for any edge-coloured odd hole, does not provide a way to compute their quantum maxima in Bell scenarios.

\begin{theorem} 
\label{thm:MaxBellCycles}
In bipartite Bell scenarios, the maximum quantum value for a graph of exclusivity $C_n$, with odd $n\ge 5$, is
\begin{equation}
MB(C_n)=\frac{1}{2}+\frac{(n-1)}{4}\left[1+\cos{\left(\frac{\pi}{n-1}\right)}\right].
\label{qb}
\end{equation}
This maximum occurs in the $(2,(n-1)/2,2)$ Bell scenario and corresponds to the maximum quantum violation of the chained Bell inequality with $(n-1)/2$ measurements per party \cite{Pearle70,BC90}.
\end{theorem}
This contrasts with the well known result \cite{AQBTC13}

\begin{equation}
\vartheta(C_n)=\frac{n \cos\left(\frac{\pi}{n}\right)}{1+\cos\left(\frac{\pi}{n}\right)}.
\label{eq:MaxHoles}
\end{equation}

%%%%%%%%%%%%%%%%%%%%%%%%%%%%%%%%%%%%%%%%%%%%%%%%%%%%%%%%%%%%%%%%%%%

\subsection{Quantum maxima for antiholes in Bell scenarios}

%%%%%%%%%%%%%%%%%%%%%%%%%%%%%%%%%%%%%%%%%%%%%%%%%%%%%%%%%%%%%%%%%%%

Finally, we now discuss the question of whether the quantum maxima of odd antiholes can be achieved in Bell scenarios. Analogously to Corollary \ref{thm:NoMaxBellHole}, we believe in a similar result for odd antiholes, expressed as the following conjecture.

\begin{conjecture}
\label{thm:NoMaxBellAntihole}
An odd antihole (i.e., the complement of an odd hole) cannot reach its quantum maximum in any Bell scenario.
\end{conjecture}

That is, Corollary \ref{thm:NoMaxBellHole} alongside with the above conjecture states that the basic structures responsible for the quantum versus classical difference, odd holes and antiholes, cannot reach their maximal quantum values when embedded into Bell scenarios.

We know, however, that an exact analogous of Theorem \ref{thm:NocolouredCyle} does not hold for odd antiholes, as we have found an edge-colouring of $\bar{C}_7$, shown in Fig.~\ref{seven} (a), which has an OPR achieving the Lovász number of $\bar{C}_7$. It turns out, however, that this colouring cannot represent the events of a Bell scenario, according to Theorem \ref{thm:BellStructure}. We thus believe in the following conjecture.

\begin{conjecture}
\label{thm:NoMaxBellAntiholecolours}
There is no orthogonal projective representation of any Bell edge-coloured odd antihole $\bar{\textsf{C}}_n$ which is a Lovász-optimal representation of the shadow $\bar{C}_n$.
\end{conjecture}

Corollary ~\ref{thm:NoMaxBellHole} follows from Theorem~\ref{thm:NocolouredCyle}, as well as Conjecture~\ref{thm:NoMaxBellAntihole} follows from Conjecture~\ref{thm:NoMaxBellAntiholecolours}. In the simplest case, $n = 5$, Conjecture \ref{thm:NoMaxBellAntiholecolours} comes from Theorem~\ref{thm:NocolouredCyle} and the fact that $C_5$ is self-complementary. We thus prove it for the first important case, $n=7$.

\begin{theorem} \label{thm:NoBellAnti7}
$MB(\bar{C}_7) < \vartheta(\bar{C}_7)$.
\end{theorem}

The proof of this theorem will not be constructive.
We actually do not know what is the maximum obtained for $\overline{C_7}$ for Bell scenarios.
On the other hand, the maximum that can be obtained without extra constraints in quantum theory is well known \cite{CDLP13}.

%%%%%%%%%%%%%%%%%%%%%%%%%%%%%%%%%%%%%%%%%%%%%%%%%%%%%%%%%%%%%%%%%%%

\section{Discussion}
\label{conclusions}

%%%%%%%%%%%%%%%%%%%%%%%%%%%%%%%%%%%%%%%%%%%%%%%%%%%%%%%%%%%%%%%%%%%

How are our results related to the problem of finding the principle of nonlocal quantum correlations? In a nutshell, our results show how, even at the level of contextuality building blocks, the constraints inherent to any Bell scenario transform quantum sets that are easy to characterise (and follow from a simple principle) into quantum sets that are difficult even to bound.

The set of quantum probability assignments for a graph of exclusivity $G$ is the set of probabilities for events having $G$ as graph of exclusivity, without specifying which scenario produces the events (all scenarios are valid). It can be proven \cite{CSW14} that, for any graph $G$, the set of quantum probability assignments is the theta body of $G$ \cite{GLS86}, which, for any $G$, is a closed set. Any element of the theta body of $G$ can be achieved in the KS contextuality scenario consisting of dichotomic observables with a graph of compatibility isomorphic to $G$ \cite{CDLP13}. Moreover, given $G$, deciding whether a probability assignment is quantum or not is the solution of a single semi-definite program \cite{GLS86}. More importantly, there is a simple principle that, for any graph, selects the theta body \cite{Cabello19a}.

In contrast to that, the set of quantum correlations (a correlation is a probability assignment for all the events produced in a scenario) for a given (Bell or KS) scenario is, in general, not closed \cite{Slofstra2019}, and the problem of whether a correlation is quantum or not is, in general, undecidable \cite{Slofstra2019}. Moreover, the maximal quantum violation of a Bell inequality may not be computable. In fact, it may be even impossible to approximate \cite{J17, JNVWY20}. All these features help us to understand why it is difficult to find a single principle that selects the quantum sets of correlations for Bell scenarios. These sets are much more difficult to characterise. Consequently, it is reasonable to expect that they are more difficult to understand.

Interestingly, the set of quantum correlations for any scenario is a subset of the set of quantum probability assignments for the graph of exclusivity of all events in that scenario. Therefore, all difficulties arise when adding scenario constraints. That is why, for identifying the principle of quantum nonlocal correlations, it is useful to start by identifying the principle of quantum probability assignments for the graphs of exclusivity \cite{Cabello19a}. But it is also important to study the transition between the respective theta bodies and the corresponding sets of correlations and identify exactly where the difference between the sets begins.

The results presented in this paper show that the differences (and the difficulties) already appear at the level of the simplest graphs of exclusivity. We have shown that, already there, the constraints of any Bell scenario exclude maximal quantum assignments.

On the one hand, this stresses the importance of experimentally testing quantum maximally KS contextual correlations for holes and antiholes. They are fundamental predictions of quantum theory that no Bell test can target. For odd holes, beautiful experiments with repeatable and minimally disturbing measurements have been performed on ions \cite{MZLCAH18}. However, for odd antiholes, so far, there are only photonic tests simulating sequential measurements \cite{ACGBXLDBSC15}. It would be interesting to test the quantum maximum for antihole correlations with repeatable and minimally disturbing measurements. One difficulty of these experiments is that $k$ antihole correlations require quantum systems of dimension $k-2$ and sequences of $(k-1)/2$ measurements \cite{CDLP13}.

On the other hand, our results leave some open questions whose answer may help us to better understand quantum theory and identify further applications. For example, 
\begin{itemize}
\item[(i)] Is there a way to prove that, in Bell scenarios, the maximum quantum value for $\overline{C_n}$, with odd $n\ge 9$, is strictly smaller than the maximum quantum value for $\overline{C_n}$?
\item[(ii)] What is the quantum maxima for odd antiholes in Bell scenarios? In which Bell scenarios do these maxima occur?
\item[(iii)] Can we achieve the quantum maxima for odd holes and antiholes in hybrid scenarios where sequential local measurements are performed on parts of an entangled system \cite{KCK14,TRC19,XXRMTCR23}? And, if so, which tasks may take advantage of these correlations?
\end{itemize}
Further research is needed to answer all these questions.

%%%%%%%%%%%%%%%%%%%%%%%%%%%%%%%%%%%%%%%%%%%%%%%%%%%%%%%%%%%%%%%%%%%

\section{Proofs}
\label{proofs}

%%%%%%%%%%%%%%%%%%%%%%%%%%%%%%%%%%%%%%%%%%%%%%%%%%%%%%%%%%%%%%%%%%%

In this section, we present the proofs of the results stated in Sec.~\ref{results}. In some of them we will use a concept which is the analogous of the Lov\'asz number for edge-coloured multigraphs.

\begin{definition} \cite{RDLTC14}
The {\em factor-constrained Lov\'asz number} of a multigraph $\textsf{G}$ composed of simple graphs $G_A = (V, E_A),\ldots, G_N = (V, E_N)$ is
\begin{equation}
\vartheta_c(\textsf{G}):= \sup \sum_{i\in V} \langle \psi | \Pi_i | \psi \rangle,
\label{mLovasz}
\end{equation}
where the supremum in the sum is taken over all orthogonal projective representations of $\textsf{G}$, unit vectors $|\psi \rangle$ in the $D$-dimensional Hilbert space of the OPR (not necessarily product vectors), and dimensions $D$ (not necessarily finite). 
\label{def:colored_lovasz}
\end{definition}

We say that an orthogonal projective representation of $\textsf{G}$ is {\em Lovász-optimal} if it realises $\vartheta_c(\textsf{G})$ for some handle. For simplicity, we will refer to $\vartheta_c(\textsf{G})$ as the {\em coloured Lov\'asz number} of $\textsf{G}$. 

%%%%%%%%%%%%%%%%%%%%%%%%%%%%%%%%%%%%%%%%%%%%%%%%%%%%%%%%%%%%%%%%%%%

\subsection{Proof of Theorem \ref{thm:NocolouredCyle}}

%%%%%%%%%%%%%%%%%%%%%%%%%%%%%%%%%%%%%%%%%%%%%%%%%%%%%%%%%%%%%%%%%%%

To prove Theorem \ref{thm:NocolouredCyle}, our strategy will be to show that, for any edge-coloured $\textsf{C}_n$ with odd $n\geq 5$ and containing at least two colours, no Lov\'asz-optimal OPR satisfies a symmetry property satisfied by all Lov\'asz-optimal OPRs of the shadow $C_n$, given by the following lemma.

\begin{lemma}\cite{Bharti2019}
\label{lem:self-test}
Consider a Lov\'asz-optimal OPR of $C_n$, with odd $n$, and handle $|\psi\rangle$ such that $\vartheta(C_n) = \sum_i \langle \psi | \Pi_i | \psi \rangle$, where $\Pi_i$ is the projector associated to vertex $i$ in this OPR. Then, for every vertex $i$, $p_n = \langle \psi|\Pi_i |\psi \rangle = \frac{cos(\pi/n)}{1+cos(\pi/n)}$. 
\end{lemma}

%%%%%%%%%%%%%%%%%%%%%%%%%%%%%%%%%%%%%%%%%%%%%%%%%%%%%%%%%%%%%%%%%%%
% Fig. 2
%%%%%%%%%%%%%%%%%%%%%%%%%%%%%%%%%%%%%%%%%%%%%%%%%%%%%%%%%%%%%%%%%%%

\begin{figure}[h]
\centering
\includegraphics[width = 8.4cm]{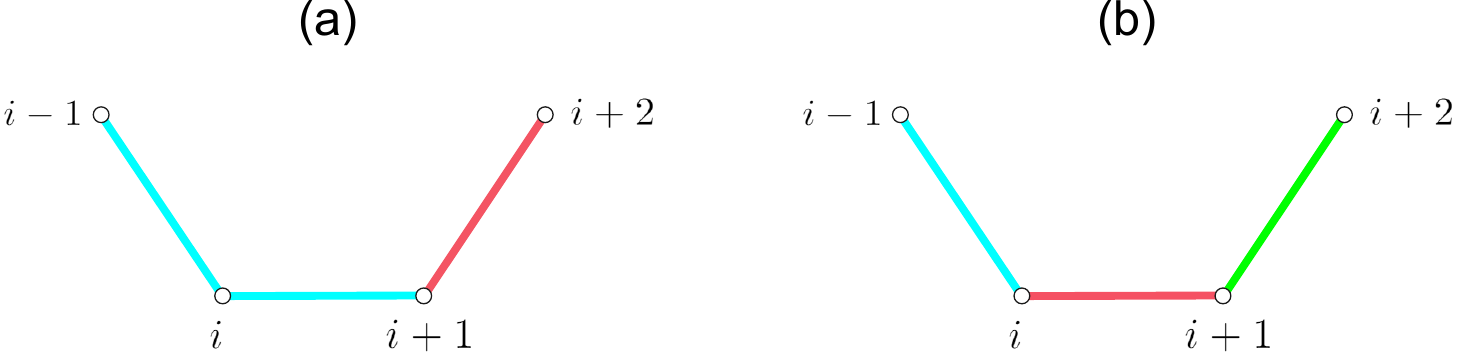}
\caption{Induced subgraphs of edge-coloured cycles used to prove Lemma~\ref{lem:size_two_propagator}, (a), and Lemma~\ref{lem:simple_propagator}, (b).}
\label{fig:proof_theo10}
\end{figure}

%%%%%%%%%%%%%%%%%%%%%%%%%%%%%%%%%%%%%%%%%%%%%%%%%%%%%%%%%%%%%%%%%%%

Now, to prove Theorem~\ref{thm:NocolouredCyle} in full generality, we would need, in principle, to deal with all possible edge-colourings of $C_n$. However, the following lemma greatly simplifies the colourings that we have to consider.

\begin{lemma} (Removing edges does not decrease the coloured Lov\'asz number)
\label{lem:NoMultipleEdges}
If $\textsf{G}$ is a coloured multigraph and $\textsf{H}$ is the coloured multigraph obtained from $\textsf{G}$ by removing an edge, then $\vartheta_c\left( \textsf{H}\right) \geq \vartheta_c\left(\textsf{G}\right)$.
\end{lemma}

{\em Proof:} By definition, $\vartheta_c\left(\textsf{G}\right)$ is a maximisation of a sum over the vertices of $\textsf{G}$, with restrictions given by the edges of $\textsf{G}$. Since $\textsf{H}$ has the same vertices as $\textsf{G}$, $\vartheta_c\left(\textsf{H}\right)$ is a maximisation of the same sum, with less restrictions (all edges of $\textsf{G}$ except for the removed edge).
\endproof 

Lemma~\ref{lem:NoMultipleEdges} implies that, among all possible colourings of $C_n$, the one with largest coloured Lovász number is one without multiple edges. Therefore, it suffices to prove Theorem~\ref{thm:NocolouredCyle} for the edge-coloured versions of $C_n$ containing only simple edges.

Notice that, for any coloured $\textsf{C}_n$ with this property, that is, without multiple edges, each vertex is only connected to two edges. Thus, in any Lov\'asz-optimal OPR of these graphs, the projector associated to each vertex $i$ can be written as $\Pi_i = \Pi_i^A \otimes \mathbb{I}^B \otimes \mathbb{I}^C \otimes \cdots \otimes \Pi_i^J \otimes \mathbb{I}^K \otimes \cdots \otimes \mathbb{I}^N = \Pi_i^A \otimes \Pi_i^J$, where $A$ and $J$ are the parties associated to the colours of the edges vertex $i$ is connected with.

Moreover, whenever neighbouring edges have different colours, we might further simplify the projectors. Suppose, for instance, that vertex $i-1$ and $i$ are connected by an $A$ coloured edge, and vertex $i$ is connected to $i+1$ by an edge of a different colour. Then, for any Lov\'asz-optimal OPR, we can write $\Pi_{i}^A = \mathbb{I}^A - \Pi_{i-1}^A$.
With these properties, we can prove the following lemmas.

\begin{lemma}
\label{lem:size_two_propagator}
Consider an edge-coloured $\textsf{C}_n$, with odd $n$, containing only simple edges and such that vertex $i-1$ is connected to vertex $i$ with an edge of colour $A$, vertex $i$ is connected to vertex $i+1$ with an edge of the same colour $A$, and the edge connecting $i+1$ and $i+2$ has a different colour $B$. See Fig.~\ref{fig:proof_theo10} (a). If a Lov\'asz-optimal OPR of this coloured $C_n$ achieves its colourless Lov\'asz number with handle $|\psi\rangle$, then, in this OPR, vertices $i$ and $i+1$ are associated to projectors $\Pi_{i} = \Pi_{i}^A \otimes \mathbb{I}$ and $\Pi_{i+1} = (\mathbb{I}^A - \Pi_i^A) \otimes \Pi_{i+1}^B$ satisfying $\langle \psi | \Pi_i^A \otimes \Pi_{i+1}^B | \psi \rangle = 0$.
\end{lemma}

{\em Proof:} Without loss of generality, we can assume that in a Lov\'asz-optimal OPR of a coloured $\textsf{C}_n$ such as described above, the projectors associated to vertices $i-1$, $i$, $i+1$, and $i+2$ can be written as $\Pi_{i-1} = \Pi_{i-1}^A \otimes \Pi_{i-1}^{B\ldots N}$, $\Pi_{i} = \Pi_{i}^A \otimes \mathbb{I}$, $\Pi_{i+1} = (\mathbb{I}^A - \Pi_i^A) \otimes \Pi_{i+1}^B$, and $\Pi_{i+2} = (\mathbb{I}^B - \Pi_{i+1}^B) \otimes \Pi_{i+2}^{AC\ldots N}$, respectively, where $\Pi_{i-1}^A \Pi_i^A = 0$. If this representation achieves the colourless Lov\'asz number of $C_n$, then we also know by Lemma~\ref{lem:self-test} that, for every vertex $i$, its associated projector satisfies $\langle \psi | \Pi_i |\psi\rangle = p_n$.

Now, consider a similar projective representation of this graph, in which we only substitute $\Pi_i$ and $\Pi_{i+1}$ for, respectively, $\Pi_i' = \Pi_{i}^A \otimes (\mathbb{I}^B - \Pi_{i+1}^B)$ and $\Pi_{i+1}' = \mathbb{I} \otimes \Pi_{i+1}^B$, while keeping the rest of the representation unchanged. This is not exactly an OPR of the considered edge-coloured graph, but notice that the projectors $\Pi_{i-1}$, $\Pi_{i}'$, $\Pi_{i+1}'$, and $\Pi_{i+2}$ satisfy $\Pi_{i-1}\Pi_i' = \Pi_i'\Pi_{i+1}' = \Pi_{i+1}'\Pi_{i+2} = 0$. That is, this new representation is an OPR of the shadow $C_n$. Moreover, since $\Pi_{i}' + \Pi_{i+1}' = \Pi_i + \Pi_{i+1}$, this OPR is Lov\'asz-optimal. Therefore, Lemma~\ref{lem:self-test} guarantees that $\langle \psi | \Pi_i^A \otimes (\mathbb{I}^B - \Pi_{i+1}^B) | \psi \rangle = \langle \psi | \Pi_i^A \otimes \mathbb{I} | \psi \rangle$, which implies $\langle \psi | \Pi_i^A \otimes \Pi_{i+1}^B | \psi \rangle = 0$.
\endproof

\begin{lemma}
\label{lem:simple_propagator}
Consider an edge-coloured $\textsf{C}_n$, with odd $n$, containing only simple edges and such that vertex $i-1$ is connected to vertex $i$ with an $A$ coloured edge, vertex $i$ is connected to $i+1$ with an edge of a different colour $B$, and vertex $i+1$ is connected to vertex $i+2$ with a $C$ coloured edge, which is a different colour than $B$, but might be the same as $A$. See Fig.~\ref{fig:proof_theo10} (b). Moreover, consider a Lov\'asz-optimal OPR with projectors $\Pi_{i-1} = \Pi_{i-1}^A \otimes \Pi_{i-1}^{B\ldots N}$, $\Pi_i = (\mathbb{I}^A - \Pi_{i-1}^A) \otimes \Pi_{i}^B$, and handle $|\psi\rangle$, satisfying $\langle \psi | \Pi_{i-1}^A \otimes \Pi_i^B | \psi \rangle = 0$ and $\langle \psi | \Pi_{i-1}^A \otimes \mathbb{I}| \psi \rangle = p_n$. Then, if this representation achieves the colourless Lovász number of $C_n$, the projector $\Pi_{i+1} = (\mathbb{I}^B - \Pi_{i}^B) \otimes \Pi_{i+1}^C$ is such that $\langle \psi | \Pi_i^B \otimes \Pi_{i+1}^C | \psi \rangle = 0$.
\end{lemma}

{\em Proof:} The proof is completely analogous to the proof of Lemma~\ref{lem:size_two_propagator}. First, notice that we can substitute $\Pi_i = (\mathbb{I}^A - \Pi_{i-1}^A) \otimes \Pi_{i}^B$ for $\Pi_{i}' = \mathbb{I} \otimes \Pi_i^B$, and then jointly substitute $\Pi_{i}'$ and $\Pi_{i+1} = (\mathbb{I}^B - \Pi_{i}^B) \otimes \Pi_{i+1}^C$ for $\Pi_{i}'' = \Pi_i^B \otimes (\mathbb{I}^C - \Pi_{i+1}^C)$ and $\Pi_{i+1}'' = \mathbb{I}\otimes \Pi_{i+1}^C$, respectively. These changes do not alter the sum $\sum_i \langle \psi | \Pi_i | \psi \rangle$, so we still get a Lov\'asz-optimal OPR of the shadow $C_n$ with the same construction used in Lemma~\ref{lem:size_two_propagator}. Then, by using Lemma~\ref{lem:self-test}, one can obtain $\langle \psi | \Pi_i^B \otimes \Pi_{i+1}^C | \psi \rangle = 0$.
\endproof

These lemmas tell us how to propagate projectors along a cycle. Suppose that there exists an OPR of a coloured version of $C_n$, with odd $n$ and without multiple edges, achieving its colourless Lov\'asz number. Moreover, suppose that in this graph there exists a path of at least size two which ends, say, in vertex $2$. Let $A$ be the colour of this path and $B$ be the colour of the edge connecting vertices $2$ and $3$. Then, we know that $\Pi_1 = \Pi_{1}^A$, $\Pi_2 = (\mathbb{I} - \Pi_1^A) \otimes \Pi_2^B$, and Lemma~\ref{lem:size_two_propagator} guarantees that $\langle \psi |\Pi_{1}^A \otimes \Pi_2^B | \psi \rangle = 0$.

If the edge connecting vertices $3$ and $4$ also has the colour $B$, then we can write $\Pi_3 = \mathbb{I} - \Pi_2^B$. However, since this is an OPR achieving the colourless Lovász number of $C_n$, by Lemma~\ref{lem:self-test} we have $p_n = \langle \psi | \Pi_2^B| \psi \rangle = \langle \psi |(\mathbb{I} - \Pi_2^B)| \psi \rangle = 1- p_n$, that is, $p_n = 1/2$. This is in contradiction with the fact that $p_n = \frac{cos(\pi/n)}{1+cos(\pi/n)} < 1/2$.

On the other hand, if the edge connecting $3$ and $4$ is of a different colour than $B$, then we can successively apply Lemma~\ref{lem:simple_propagator} until we find a path of at least size two, in which, once more, the contradiction described above occurs. In the worst case, this path will be the one we started with.

The proof is almost complete, but there is a gap yet to be filled, which is the fact that not all coloured $\textsf{C}_n$'s have a path of at least size two, a necessary ingredient for the above reasoning. However, the following lemma says that, if this path does not exist, we can construct one. 

%%%%%%%%%%%%%%%%%%%%%%%%%%%%%%%%%%%%%%%%%%%%%%%%%%%%%%%%%%%%%%%%%%%
% Fig. 3
%%%%%%%%%%%%%%%%%%%%%%%%%%%%%%%%%%%%%%%%%%%%%%%%%%%%%%%%%%%%%%%%%%%

\begin{figure}
\hspace{-2mm}
\includegraphics[width=8.6cm]{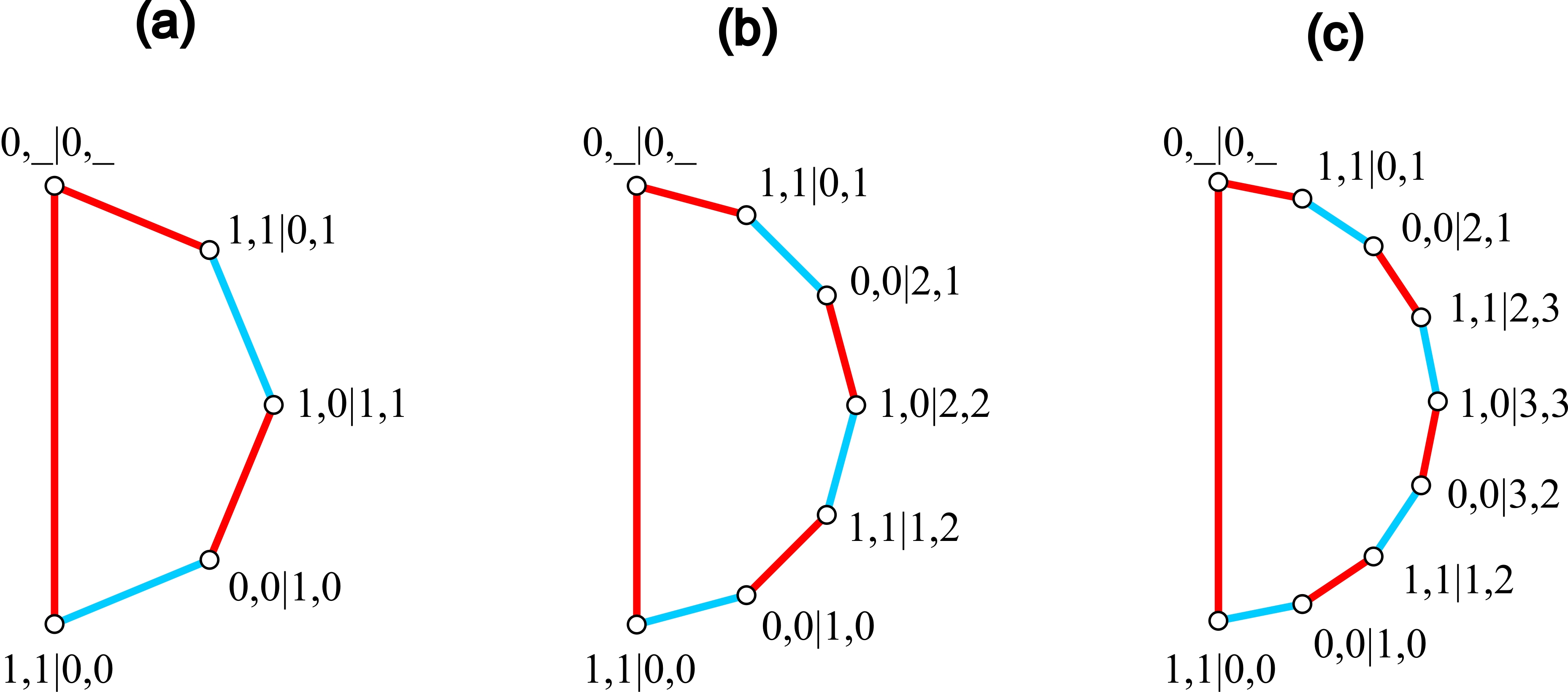}
\caption{Coloured multigraphs corresponding to events in the Bell scenario $\left(2,{(n-1)}/{2},2\right)$ and providing the maximum quantum Bell nonlocal value for holes. $a,b|x,y$ denotes the event in which Alice measures $x$ and obtains $a$, and Bob measures $y$ and obtains $b$. The ``\_'' on the labelings of some events means that the respective measurement is ignored.}
\label{Fig0}
\end{figure}

%%%%%%%%%%%%%%%%%%%%%%%%%%%%%%%%%%%%%%%%%%%%%%%%%%%%%%%%%%%%%%%%%%%

\begin{lemma}(Merging colours)
\label{lem:Mergingcolours}
If $\textsf{G}$ is a coloured graph with $q$ colours, and $\textsf{H}$ is the $\left(q-1\right)$-coloured graph obtained from $\textsf{G}$ by merging two colours, then $\vartheta_c\left(\textsf{H}\right) \geq \vartheta_c\left(\textsf{G}\right)$. 
\end{lemma}

{\em Proof:} Any valid OPR for $\textsf{G}$ induces an OPR for $\textsf{H}$ when we consider for the merged colours the tensor product of their vector spaces as just one factor.
\endproof

If a coloured $\textsf{C}_n$, with $n$ odd and without multiple edges, has only two colours, then there always exists a path of at least size two. If there are more colours and there is no such path, then Lemma~\ref{lem:Mergingcolours} allows us to simply merge two of the colours in order to construct it. This concludes the proof of Theorem~\ref{thm:NocolouredCyle}. 

%%%%%%%%%%%%%%%%%%%%%%%%%%%%%%%%%%%%%%%%%%%%%%%%%%%%%%%%%%%%%%%%%%%

\subsection{Proof of Theorem \ref{thm:BellStructure}}

%%%%%%%%%%%%%%%%%%%%%%%%%%%%%%%%%%%%%%%%%%%%%%%%%%%%%%%%%%%%%%%%%%%

Let us start from a {given} Bell scenario.
A Bell {inequality can be} written as a weighted sum of probabilities of events $a_I|A_I$, where $I$ is the list of parts making measurements for this event, $A_I$ is the ordered list of measurements chosen for each part, and $a_I$ is the corresponding list of outputs.
Those events correspond to the vertices of the coloured graph.
Two vertices $a_I|A_I$ and $b_J|B_J$ are linked in the colour $c$ corresponding to some part whenever this part $c$ appears in $I$ and $J$, with $A_c = B_c$, while $a_c \neq b_c$.
Let us now focus on the simple graph of colour $c$.
Each measurement $M$ in part $c$ gives rise to one piece (component) of this graph (notice that a measurement will only appear as a component of the graph if at least two different outputs appear in different events in the inequality).
Events with different outputs for the same measurement are always connected. 
Since there can be different events with the same output for $M$ in part $c$, this says that this piece is not necessarily a complete graph, but characterises it as a complete $o_{M}$-partite graph, where $o_M$ is the number of different outputs for $M$ appearing at this inequality.
This proves one side of the assertion.

The other side of the assertion was explained and essentially proved in Sect.~\ref{results}.
Given a coloured graph in which each colour $c$ is the disjoint union of $m_c$ components, each of them an $o_{j_c}$-complete graph, each vertex will be labelled as $a_I|A_I$, where $I$ lists the colours for which there is some edge incident on it, and for each piece for each colour, a label is chosen for a measurement, while each independent set in this piece receives a label for an output.
\endproof 

%%%%%%%%%%%%%%%%%%%%%%%%%%%%%%%%%%%%%%%%%%%%%%%%%%%%%%%%%%%%%%%%%%%

\subsection{Proof of Theorem \ref{thm:MaxBellCycles}}

%%%%%%%%%%%%%%%%%%%%%%%%%%%%%%%%%%%%%%%%%%%%%%%%%%%%%%%%%%%%%%%%%%%

To prove Theorem~\ref{thm:MaxBellCycles}, we only need to consider two-colour edge-coloured $\textsf{C}_n$, since we are only concerned with bipartite Bell scenarios. Also, because of Lemma \ref{lem:NoMultipleEdges}, we need to consider only graphs without multiple edges.

Then, to begin with, we prove a lemma which is useful to simplify even further the graphs we have to analyse. It allows us to transform coloured cycles into simpler ones while keeping track of the changes in their coloured Lovász number.

\begin{lemma}\label{lem:PlusOne}
Consider an edge-coloured n-cycle $\textsf{C}_n$, without multiple edges, and such that the edges $\{i-1, i\}$, $\{i,i+1\}$ and $\{i+1, i+2\}$ are all of the same colour $A$, and the edges $\{i-2, i-1\}$ and $\{i+2, i+3\}$ are not of the colour $A$. Then, consider the graph $\textsf{C}_{n-2}$ which is constructed from $\textsf{C}_n$ by excluding the vertices $i$ and $i+1$ and connecting vertices $i-1$ and $i+2$ with an edge of colour $A$. Then, $\vartheta_c(\textsf{C}_n) = \vartheta_c(\textsf{C}_{n-2}) + 1$. 
\end{lemma}

{\em Proof:} Consider a Lovász-optimal OPR of the graph $\textsf{C}_n$. Without loss of generality, we can assume $\Pi_{i-1} = (\mathbb{I}^A - \Pi_{i}^A) \otimes \Pi_{i-1}^B$, $\Pi_i = \Pi_i^A$, $\Pi_{i+1} = \Pi_{i+1}^A$ and $\Pi_{i+2} = (\mathbb{I}^A - \Pi_{i+1}^A) \otimes \Pi_{i+2}^C$. However, we know that $\Pi_{i+1}^A \leq \mathbb{I}^A - \Pi_{i}^A$ and, moreover, for whatever projectors $\Pi^A$ and $\Pi^C$, it holds that $\Pi^A \otimes \mathbb{I}^C \geq \Pi^A \otimes \Pi^C$. Therefore, we can take $\Pi_{i+1}^A = \mathbb{I}^A - \Pi_{i}^A$. This leads to an OPR whose restriction to all the vertices of $\textsf{C}_n$, except $i$ and $i+1$, is also an OPR of the graph $\textsf{C}_{n-2}$. Now, given that $\Pi_i + \Pi_{i+1} = \mathbb{I}$, it follows that $\vartheta_c(\textsf{C}_n) = \vartheta_c(\textsf{C}_{n-2}) + 1$. \endproof

We can combine this Lemma with the ideas used to proof Lemma~\ref{lem:size_two_propagator} to prove the following.

\begin{lemma}\label{lem:BreakingLemma}
Consider two coloured cycles $\textsf{C}_n$ and $\textsf{C}_{n-2}$, without multiple edges, such that $\textsf{C}_n$ is constructed from $\textsf{C}_{n-2}$ by creating a vertex $i'$ between the vertices $i$ and $i+1$ of $\textsf{C}_{n-2}$, creating two new edges $\{i, i'\}$ and $\{i', i+1\}$ of the same colour of the edge $\{i, i+1\}$, and removing this edge. To complete the construction of $\textsf{C}_n$, this procedure is repeated in any other edge $\{j, j+1\}$ of $\textsf{C}_{n-2}$. Then, $\vartheta_c(\textsf{C}_n) = \vartheta_c(\textsf{C}_{n-2}) + 1$.
\end{lemma}

{\em Proof:} Notice that the idea of changing the OPR in a size two path just like described in the proof of Lemma \ref{lem:size_two_propagator} is equivalent to changing the colouring of the cycle in such a way as to move the size two path along the graph. One can then take $\textsf{C}_n$ into a different edge-coloured graph which is related to $\textsf{C}_{n-2}$ in the same way as described on Lemma \ref{lem:PlusOne}. \endproof

Now, taking into account Theorem \ref{thm:BellStructure}, a Bell colouring of a cycle cannot have a path of size three or larger. Thus, a Bell colouring of a cycle only contains paths of size one or two. Moreover, given that we are restricted to two colours, an odd cycle must contain an odd number of paths of size two (as well as an odd number of paths of size one).

So, in the first place, consider an odd edge-coloured hole $\textsf{C}_n$ with only one size-two path. 
One can verify that the only event assignments consistent with this exclusivity structure gives rise to an inequality which is a re-scaled and shifted chained inequality. Fig.~\ref{Fig0} shows three coloured multigraphs corresponding to events in the Bell scenario $\left(2,{(n-1)}/{2},2\right)$ \cite{Pearle70,BC90,AQBTC13}. The quantum maximum of a chained inequality is well known \cite{Wehner2005}, and we can use it to conclude that 
\begin{equation}
\vartheta_c(\textsf{C}_n)=\frac{1}{2}+\frac{(n-1)}{4}\left[1+\cos{\left(\frac{\pi}{n-1}\right)}\right].
\label{qb}
\end{equation}
More generally, if $\textsf{C}_n$ has $t$ paths of size two, it is possible to use Lemma \ref{lem:BreakingLemma} to reduce it to an odd hole with only one path of size two, while taking into account the unit changes in the coloured Lovász number. Thus, in this case,
\begin{equation}
\vartheta_c(\textsf{C}_n)= \frac{t}{2}+\frac{n-t}{4} \left[1+\cos \left( \frac{\pi}{n-t} \right) \right].
\end{equation}
The maximum of this expression is attained when $t = 1$, and thus the proof is concluded.

%%%%%%%%%%%%%%%%%%%%%%%%%%%%%%%%%%%%%%%%%%%%%%%%%%%%%%%%%%%%%%%%%%%
% Fig. 4
%%%%%%%%%%%%%%%%%%%%%%%%%%%%%%%%%%%%%%%%%%%%%%%%%%%%%%%%%%%%%%%%%%%

\begin{figure}
\hspace{-2mm}
\includegraphics[width=8.6cm]{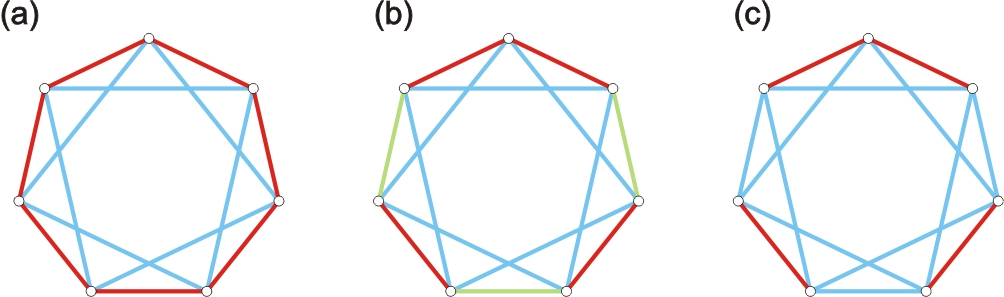}
\caption{Coloured multigraphs used in the proof of Theorem~\ref{thm:NoBellAnti7}.}
\label{seven}
\end{figure}

%%%%%%%%%%%%%%%%%%%%%%%%%%%%%%%%%%%%%%%%%%%%%%%%%%%%%%%%%%%%%%%%%%%

\subsection{Proof of Theorem \ref{thm:NoBellAnti7}} 

%%%%%%%%%%%%%%%%%%%%%%%%%%%%%%%%%%%%%%%%%%%%%%%%%%%%%%%%%%%%%%%%%%%

We begin by computing the factor-constrained Lov\'asz number for all possible ways of colouring $\overline{C_7}$ with two colours. 
We observe that there is only one way such that $\vartheta_c(\overline{\textsf{C}_7}) = \vartheta (\overline{C_7})$; this colouration is presented in Fig.~\ref{seven} (a), where $\overline{\textsf{C}_{7}}$ is `factored' into two $C_{7}$'s, coloured distinctively (note that this colouring does not represent a set of events of a Bell scenario).
Suppose that one wants to try to obtain this value in a Bell scenario. 
The merging colours Lemma~\ref{lem:Mergingcolours} implies that it would be necessary to obtain this specific two-colouring whenever merging colours into two sets.
Let us focus just on one factor.
As a $C_7$, it can be Bell-coloured in many ways, for example, as in Fig.~\ref{seven} (b).
However, this allows for other colour merging, merging some of the new colours to the one of the other previous $C_7$ factor, like in Fig.~\ref{seven} (c).
This contradiction shows that any Bell coloured $\overline{C_7}$ will necessarily have $\vartheta_c (\overline{\textsf{C}_7}) < \vartheta (\overline{C_7})$.

Another way of phrasing the above proof is that, since the only bi-colouring that reaches $\vartheta (\overline{C_7})$ is not Bell, more colours will be needed. 
By merging colours, different upper bounds will be given by each bi-coloured $\overline{C_7}$ obtained in the process. 
Naturally, in the end, the smaller upper bound over all merging colours processes is the relevant one.
Since every other bi-colouring implies $\vartheta_c(\overline{\textsf{C}_7}) < \vartheta (\overline{C}_7)$, no Bell colouring of $\overline{\textsf{C}_7}$ can reach $\vartheta(\overline{C}_7)$.

%%%%%%%%%%%%%%%%%%%%%%%%%%%%%%%%%%%%%%%%%%%%%%%%%%%%%%%%%%%%%%%%%%%

\section*{Acknowledgments}

%%%%%%%%%%%%%%%%%%%%%%%%%%%%%%%%%%%%%%%%%%%%%%%%%%%%%%%%%%%%%%%%%%%

We thank J.\ R.\ Portillo for useful conversations. 
This work was supported by the program Science without Borders (\href{http://dx.doi.org/10.13039/501100003329}{CAPES} and \href{http://dx.doi.org/10.13039/501100003593}{CNPq}), 
by the S\~{a}o Paulo Research Foundation FAPESP (grants nos.\ 2018/07258-7 and 2023/04197-5), by {\textit{C\'atedras Ibero-Americanas}} (Banco Santander, Call 062/2017), by CNPq under grant no.\ 310269/2019-9, by the QuantERA grant SECRET, by \href{10.13039/501100011033}{MCINN/AEI} (project no.\ PCI2019-111885-2) and by \href{10.13039/501100011033}{MCINN/AEI} (project no.\ PID2020-113738GB-I00).

%%%%%%%%%%%%%%%%%%%%%%%%%%%%%%%%%%%%%%%%%%%%%%%%%%%%%%%%%%%%%%%%%%%

%%%%%%%%%%%%%%%%%%%%%%%%%%%%%%%%%%%%%%%%%%%%%%%%%%%%%%%%%%%%%%%%%%%

\end{document}